\documentclass[a4paper]{jpconf}

\usepackage{graphicx}

\begin{document}


\title{Experimental approaches for 100~TeV astronomy}

\author{Pierre Colin$^1$, Stephan LeBohec$^1$ and Jamie Holder$^2$}

\address{$^1$Department of Physics, University of Utah, Salt Lake City, Utah, USA}

\address{$^2$Department of Physics and Astronomy, University of Delaware, Newark, Delaware, USA}

\ead{colin@physics.utah.edu}

\begin{abstract}
The high energy end of $\gamma$-ray source spectra might provide
important clues regarding the nature of the processes involved in
$\gamma$-ray emission. Several galactic sources with hard emission spectra 
extending up to more than 30~TeV have already been reported. Measurements 
around 100~TeV and above should be an important goal for the next generation 
of high energy $\gamma$-ray astronomy experiments. Here we present several 
techniques providing the required exposure ($\sim100~km^2\cdot h$). We focus 
our study on three Imaging Atmospheric Cherenkov Technique (IACT) based 
approaches: low elevation observations, large field of view telescopes, and
large telescope arrays. We comment on the advantages and disadvantages
of each approach and report simulation based estimates of their energy ranges and sensitivities.
\end{abstract}

\section{Introduction}

Questions about the origin of galactic Cosmic Rays (CR) have
motivated the development of $\gamma$-ray astronomy over 100~GeV
as $\gamma$-rays from $\pi^0$ decay were expected to
trace CR acceleration sites~\cite{Drury}. However, measured TeV spectra are 
most generally explained in terms of Inverse Compton (IC) scattering of photons  
by high energy electrons. Current observations do not allow 
to unambiguously identify a hadron component in $\gamma$-ray spectra
\cite{Magic}. At higher energy, synchrotron cooling of 
electrons becomes more important and should cause the IC component of spectra 
to soften. A cutoff is expected below or around 100~TeV. The $\gamma$-rays 
resulting from hadron processes may not display a similar cutoff 
~\cite{Rowell}. Their contribution could extend up to several hundred TeV 
depending on CR acceleration maximum energy in the source. Spectral 
measurements around 100~TeV could provide clear discrimination between 
electron and hadron processes.

Current experiments are not likely to reach 100~TeV because their limited 
collection area makes the required exposures unreasonable. It required 5~years
($\sim 400~h$ of observation) with the HEGRA experiment~\cite{HEGRA_Crab} to measure 
the Crab nebula spectrum up to 80~TeV, the highest astrophysical $\gamma$-ray
energy ever reported. From extrapolating galactic source spectra measured with 
HEGRA and HESS~\cite{HESS_3}, it appears future projects need to achieve at 
least $100~km^2\cdot h$ (Figure~\ref{fig}) to obtain meaningful measurements 
at 100~TeV.

\section{Techniques to Obtain the Required Exposure}

\subsection{Shower front samplers}

Ground detectors arrays like Milagro~\cite{Milagro} detect charged particles 
in the air-shower tail. They benefit from a close to $100\%$ duty cycle and
with their large field of view they can observe each source for $\sim
1500~h/year$.
Milagro already offers a large exposure 
($\sim 40~km^2\cdot h/year$), but its sensitivity is limited by its angular 
resolution and its CR discrimination. The HAWC project~\cite{HAWC} should be 
a strong improvement over this. Its effective collection area was estimated 
up to 10~TeV and is expected to remain constant($\sim 0.07~km^2$) at higher 
energy. HAWC will reach a $100~km^2\cdot h$ exposure per year.

\subsection{Imaging Atmospheric Cherenkov Technique }

IACT can be used only during clear, and usually moonless, nights. The duty cycle is about 
10\%. Maximal observation time for a source is 200-300~h/year. In order for 
their science program to be diversified, small field of view (FOV) IACT  
generally are not dedicated to any given source for more than 100~h/year. 
To achieve $100~km^2\cdot h$ exposure per year, future IACT 
experiments need to offer effective collection areas of at least $1~km^2$ with 
a small FOV or $0.5~km^2$ with full-sky survey capability.

The ground level light density is relatively constant within the 
central plateau ($\sim 130~m$ radius for vertical showers) of the Cherenkov 
light pool and decreases rapidly with the distance from the shower core 
outside the central plateau. At large zenithal angle the plateau radius is 
much larger ($>$400m at $20^\circ$ of elevation). The three approaches we 
identified to use the Cerenkov light distribution are discussed in the next 
section.

\begin{figure}

\includegraphics[width=\columnwidth]{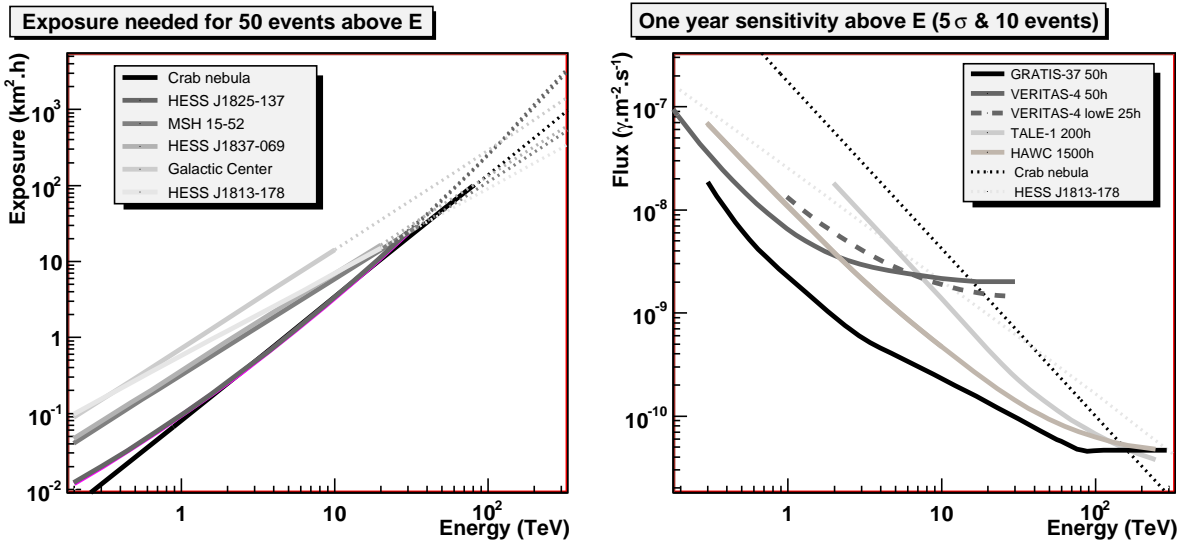}

\caption[short]{The left-hand side shows the exposure for detecting 
50 $\gamma$-rays from galactic sources as a function of the energy. The 
right-hand side shows the sensitivities of several projects (described in the 
text) compared to a few galactic source spectra.}

\label{fig}

\end{figure}

\section{Simulation of IACT approaches}

\subsection{Low elevation observation}

At low elevation showers are developing much further away and a single 
telescope can reach a $1~km^2$ collection area even with a small FOV 
($4^\circ$). This observation technique is common~\cite{Krennrich_Mkr421}, 
providing large collection areas at very high energies. However, several 
problems and difficulties appear: small images provide poor discrimination and 
shower reconstruction, large base lines are still necessary for good stereoscopic 
views, atmospheric monitoring becomes critical and the observing program is
much more constrained. For these reasons, a dedicated low elevation experiment does not 
seem very appealing. Figure~\ref{fig} shows our estimate of 
VERITAS~\cite{VERITAS} 25h sensitivity at low elevation.

\subsection{Large field of view telescopes}

In order to exploit the tail of the Cherenkov light pool at high elevation, 
IACT telescopes need a large FOV ($\sim 12^\circ$ for $1~km^2$
effective area)~\cite{delaCalle}. We simulated an $11~m^2$ telescope with a $16^\circ
\times 16^\circ$ camera based on the design of the TALE
project~\cite{TALE} which includes a "Fly's Eye" air fluorescence
telescope for the study of CR around $10^{17}~eV$. With an improved 
$0.5^\circ$ pixel camera and 20MHz FADC, the energy threshold is estimated at 
$\sim 2~TeV$ in the plateau and increase as a function  impact parameter 
($10~TeV$ at $400m$ and $\sim 100~TeV$ at $800m$). The effective
collection area increases with energy advantageously for sensitivity but 
makes spectral analysis strongly Monte-Carlo dependent. The angular
resolution is about $0.25^\circ$ at $10~TeV$ and less than $0.15^\circ$ over 
$30~TeV$. Assuming a yearly observation time of 200~h (time expected for an 
all sky-survey over $30^\circ$ elevation), we estimate the point source 
sensitivity of a TALE like experiment (see figure~\ref{fig}). This curve does 
not include the effects of $\gamma$-ray shower image shape discrimination 
capabilities which still have to be studied.

\subsection{Large Telescope arrays}

A large area telescope array (CTA, HE-Astro~\cite{HE-Astro},
TenTen~\cite{TenTen}, GRATIS) could be made to indefinitely extend the collection area.
Projects of this type are expensive and challenging because of the large number of units
involved. They also offer the best $\gamma$-ray discrimination and
reconstruction capabilities~\cite{Plyash}.
The Gamma-Ray Astrophysical Telescope Imaging System (GRATIS)
is a minimal approach to a 0.3-100~TeV large array,
consisting of 37 $5.4~m$ diameter telescopes with $4^\circ$ FOV (253 $0.25^\circ$
pixels), covering $1~km^2$ in a $200~m$ spaced hexagonal lattice. Any point 
in the array is less than 115~m from a telescope ($<$Cherenkov plateau radius).
From simulations of just one triangular cell we estimate the
energy threshold ($\sim 300~GeV$), angular resolution ($\sim 0.15^\circ$ at 
$1~TeV$ and  $\sim 0.05^\circ$ above $10~TeV$) and CR rejection. 
Figure~\ref{fig} shows the GRATIS sensitivity of the 54 triangles in the 
array. This sensitivity is conservative at high energies as showers do trigger 
more than 3 telescopes improving rejection and angular resolution, and because 
$\gamma$-rays with impact parameters outside the array are not included in this estimate.

\section{Conclusion}

Recent findings in the field of TeV astronomy provide
motivation for extending the covered energy range to more than
100~TeV where $\gamma$-ray emission models can be better discriminated.
Simulation-based estimates of various experimental approaches to 100~TeV 
astronomy are compared. Large arrays of IACT telescopes seem the most 
attractive from the point of view of the sensitivity they offer. Even a 
relatively low cost (\$ 17M) project such as GRATIS provides sufficient
sensitivity to measure many galactic source spectra from 300~GeV
to more than 100~TeV. With their all sky survey capability in an uncharted 
energy window, projects like HAWC or even TALE have a great and complementary
exploratory potential.\\


\end{document}